\documentstyle[12pt]{article}
\newcommand{\slp}{\raise.15ex\hbox{$/$}\kern-.57em\hbox{$\partial$}}
\newcommand{\sla}{\raise.15ex\hbox{$/$}\kern-.57em\hbox{$a$}}
\newcommand{\slA}{\raise.15ex\hbox{$/$}\kern-.57em\hbox{$A$}}
\newcommand{\slb}{\raise.15ex\hbox{$/$}\kern-.57em\hbox{$b$}}
\newcommand{\be}{\begin{equation}}

\newcommand{\ee}{\end{equation}}
\newcommand{\bear}{\begin{eqnarray}}
\newcommand{\ear}{\end{eqnarray}}
\newcommand{\eear}{\end{eqnarray}}
\newcommand{\ba}{\begin{eqnarray*}}
\newcommand{\ea}{\end{eqnarray*}}
\begin{document}
\begin{titlepage}
\setcounter{page}{1}
\begin{flushright}
HD--THEP--99--37\\
\end{flushright}
\vskip1.5cm
\begin{center}
{\large{\bf Recursive Construction of Generator for Lagrangian Gauge Symmetries
}}\\
\vspace{1cm}
R. Banerjee, 
%%%%%
\footnote{email: r.banerjee@thphys.uni-heidelberg.de\\
On leave of absence from S.N. Bose Natl. Ctr. for Basic Sc., Salt Lake, 
Calcutta 700091, India }
%%%%%
H.J. Rothe 
%%%%%%
\footnote{email: h.rothe@thphys.uni-heidelberg.de}
%%%%%%
and 
K. D. Rothe 
%%%%%%%
\footnote{email: k.rothe@thphys.uni-heidelberg.de}
%%%%%%
\\
{\it Institut  f\"ur Theoretische Physik - Universit\"at Heidelberg}
\\
{\it Philosophenweg 16, D-69120 Heidelberg, Germany}

{(July 27, 1999)}
\end{center}

\begin{abstract}
\noindent
We obtain, for a subclass of structure functions characterizing a 
first class Hamiltonian system,  recursive relations from which
the general form of the local symmetry transformations can be constructed in terms of the independent gauge parameters. We apply this to a
non-trivial Hamiltonian system involving two primary constraints,
as well as two secondary constraints of the Nambu-Goto type.

\end{abstract}
\end{titlepage}
%%%%%%%%%%%%%%%%%%%%%%%%%%%%%%%%%%%%%%%%%%%%%%%%%%%%%%%%%%%%%%%%%%%%%%%%%%%
%%%%%%%%%%%%%%%%%%%%%%%%%%%%%%%%%%%%%%%%%%%%%%%%%%%%%%%%%%%%%%%%%%%%%%%%%%%%

\newpage

The problem of finding the most general local symmetries of a Lagrangian
has been pursued by various authors, using either Lagrangian 
\cite{Mukunda,Gitman,Chaichian1,Shirzad1} or Hamiltonian techniques \cite{Girotti,Henneaux,Chaichian,Wipf}. 

In a recent paper \cite{Banerjee1} we had shown that the requirement of commutativity
of the time derivative operation with an arbitrary infinitesimal gauge variation generated by the first class constraints was the only input needed
for obtaining the restrictions on the gauge parameters entering the most general form of the generator of Lagrangian symmetries. The analysis was performed entirely in the
Hamiltonian framework.
On the basis of this commutativity requirement, 
we subsequently derived \cite{Banerjee2} a simple differential equation for the generator encoding, in particular, the restrictions on the gauge parameters. 

In this paper we shall obtain, for a subclass of structure functions 
characterizing a first class Hamiltonian system, the explicit solution of the above differential equation in the form of simple 
recursive relations. We then apply this general scheme to a non-trivial model discussed
in the literature \cite{Batlle}, whose secondary (first class) constraints
are identical with the primary constraints of the Nambu-Goto model. Our result for the gauge transformation is found to agree with that quoted in the literature.

We shall consider purely first class systems. The extension to mixed first and second class systems is 
straightforward. To keep the algebra simple we assume all constraints to be irreducible. 

Consider a Hamiltonian system whose dynamics is described by the
total Hamiltonian
%%%%
\footnote{We follow here the notation of Ref. \cite{Banerjee1}.}
%%%%
\be\label{totalhamiltonian}
H_T = H_c + \sum_{a_1} v^{a_1}\Phi_{a_1}\;.
\ee
where $H_c$ is the canonical Hamiltonian, $\{\Phi_{a_1}\approx 0\}$ are the 
(first class) primary constraints, and $v^{a_1}$ are the associated
Lagrange multipliers. We denote the complete set of 
(primary and secondary) constraints
%%%
\footnote{``Secondary" refers to all generations of constraints beyond the primary one.}
%%%
by
$\{\Phi_a\} = \{\Phi_{a_1},\Phi_{a_2}\}$.

Following the conjecture of Dirac \cite{Dirac}, the generator of the gauge transformations $G$ is given
by 
\be\label{generator}
G = \sum_a \epsilon^a \Phi_a
\ee
where the gauge parameters 
are allowed to depend in general on time, as well as on the 
phase space variables and Lagrange multipliers. An infinitesimal transformation on the coordinates, generated by $G$, is then given by
\be\label{delta}
\delta q^{\ell} = \epsilon^a[q^{\ell},\Phi_a]\;,
\ee
where a summation over repeated indices is always understood. 

The Poisson algebra of the constraints with themselves
and with the canonical Hamiltonian, is of the form
\be\label{algebra}
[H_c,\Phi_a] = V_a^b \Phi_b\label{algebra1}
\ee
\be
[\Phi_a,\Phi_b] = C_{ab}^c\Phi_c\label{algebra}
\ee
where $V_a^b$ and $C_{ab}^c$ may be functions of the 
phase-space variables.

As was shown in \cite{Henneaux,Banerjee1,Banerjee2}, $G$ in (\ref{generator}) will
generate a local symmetry of the corresponding total Lagrangian, provided the
following relations hold:
\bear
\delta v^{b_1}&=& \frac{d\epsilon^{b_1}}{dt}   - \epsilon^a 
[V_a^{b_1}  +  v^{a_1} C_{a_1a}^{b_1}]\;,\label{multiplier}\\
 0&=&\frac{d\epsilon^{b_2}}{dt}  - \epsilon^a 
[V_a^{b_2}  +  v^{a_1} C_{a_1a}^{b_2}]\;. \label{parameter}
\eear
In the above equations, $\frac{d\epsilon^a}{dt}$ denotes the total time derivative. For obtaining the generator of the symmetries of the original
Lagrangian, only eq. (\ref{parameter})
is relevant. Eq. (\ref{multiplier})
is required for consistency on Hamiltonian level.

As was shown in \cite{Banerjee2}, the above equations
can be compactly summarized in a simple differential equation for the
Generator $G$ expressing its time independence:
\be\label{Master}
\frac{\partial G}{\partial t} + [G,H_T] = 0\;.
\ee
Equations (\ref{multiplier}) and (\ref{parameter}) describe the 
restrictions imposed on the Lagrange multipliers and gauge parameters
for the most general case where the structure functions depend on
coordinates and momenta.
We now seek a solution of (\ref{parameter}) under two assumptions:

i) The Poisson bracket of any constraint with the primary constraints is a linear combination of only the primary constraints. 
This implies $C_{a_1a}^{b_2} = 0$, and hence the absence of the last term in (\ref{parameter}).

ii) The structure functions $V_a^{b_2}$ are constants.

These conditions appear at first sight to be very restrictive.
However, many of the physically interesting theories, fall into this class.

The problem of finding the generator of gauge transformations
subject to the above assumptions was considered in \cite{Shirzad2}.
Starting from the general set of equations (\ref{parameter}),
we present here a more compact and transparent approach to the solution,
which in fact will include the case of field dependent structure functions
$V_a^{b_2}$, if there are no terciary constraints.

In order to solve the equations (\ref{parameter}) it is convenient to 
organize the constraints into ``families", where the parent of each family
``a" is given by a primary constraint $\phi^{(a)}_0$, 
and the remaining members are recursively derived from \cite{Shirzad2}
\be\label{family}
[H_c,\phi^{(a)}_{i-1}] = \phi^{(a)}_i \;,\quad i=1,...,N_a\;.
\ee
With an obvious change in notation, this implies that the corresponding structure functions satisfy 
\be\label{structureconstants}
V_{ij}^{ab} = \delta^{ab}\delta_{i,j-1}\;,\quad i=0,\cdot\cdot\cdot ,N_a-1\;.
\ee
In order to ensure that the constraints thus obtained are irreducible,
we must adopt some systematic procedure.
A possibility is to perform this iteration level by level, for all primary 
constraints, simultaneously. We terminate a family ``a", if at a given level
$N_a$, the Poisson bracket of the constraint $\phi_{N_a}^{(a)}$ with
$H_c$ can be written as a linear combination of the members of {\it all} the
families up to this level. However, independent of the specific procedure
chosen, the Poisson bracket
of the final member of each family with $H_c$ is given by
\be\label{highestmember}
[H_c,\phi_{N_a}^{(a)}] = \sum_{b=1}^M \sum_{j=0}^{N_b} V_{N_a j}^{ab}
\phi^{(b)}_j \;,
\ee
where $M$ is the number of (independent) primary constraints.

In the new notation, equation (\ref{parameter}) reads
\be\label{parameter2}   
0=\frac{d\epsilon^{(a)}_i}{dt}  - \sum_{b=1}^M \sum_{j=0}^{N_b}\epsilon^{(b)}_j 
V_{ji}^{ba}\;, \quad i=1,\cdot\cdot\cdot,N_a\;. 
\ee
Choosing as our independent parameters the ones associated with the
last member in each family,
\be\label{indepparam}
\alpha^a := \epsilon^{(a)}_{N_a}(t)\;,
\ee
the equations (\ref{parameter2}) take the form
\be\label{parameter3}
\frac{d\epsilon^{(a)}_i}{dt} - \epsilon^{(a)}_{i-1}
-\sum_{b=1}^M \alpha^{b} V_{N_b i}^{ba}=0\;,\quad i=1,\cdot\cdot\cdot ,N_a\;. 
 \ee
The solution to this set of equations can be constructed iteratively,
by starting with the last member of a family:
\be\label{startingpoint}
\epsilon^{(a)}_{N_a-1} = \frac{d\alpha^a}{dt} - \sum_{b=1}^M \alpha^b V_{N_bN_a}^{ba}\;.
\ee
Continuing in the same fashion, one easily sees that the general solution can be written in the form
\be\label{generalsolution}
\epsilon^{(a)}_i = \sum_{n=0}^{N_a-i} \sum_{b=1}^M
\frac{d^n\alpha^b}{dt^n} A_{i(n)}^{ba}\;.
\ee
with the normalization
\be\label{normalization}
A_{N_a(0)}^{ab} = \delta^{ab}\;,
\ee
following from our choice of parametrization (\ref{indepparam}).
Substituting the above ansatz into (\ref{parameter3}) and comparing
powers in the time derivatives, we obtain the recursion relations
\be\label{Arecursion}
A_{i(n-1)}^{ab} = A_{i-1(n)}^{ab}\;,\quad i=1,\cdot\cdot\cdot,N_a,
\ee
with the ``initial conditions"
\be\label{initialcond}
A_{i-1(0)}^{ab} =  -V_{N_ai}^{ab}\;,\quad i=1,\cdot\cdot\cdot,N_a.
\ee
It is easy to see, that these recursion relations
determine the complete solution, from which the generator of the Lagrangian
gauge symmetries can be obtained. 
Using (\ref{generalsolution}) in the generator (\ref{generator}), the infinitesimal gauge transformation (\ref{delta}) takes the form
\be
\delta q^{\ell} =  \sum_{b=1}^M \sum_{n\ge 0} 
\frac{d^n\alpha^b}{dt^n}\rho^{\ell}_{(n)b}(q,\dot q) \;,
\ee
where
\be\label{Gitman}
\rho^{\ell}_{(n)b}(q,\dot q) = \sum_{a=1}^M \sum_{j\ge 0} 
 \theta(N_a-n-j)A_{j(n)}^{ba}\frac{\partial \phi^a_j}{\partial p_{\ell}}\;,
\ee
with $\theta(0)=1$, and where it is understood that the canonical momenta are to be replaced
by the respective expressions in terms of the Lagrangian variables.
Expression (\ref{Gitman}) is in the form obtained by purely
Lagrangian methods \cite{Gitman,Shirzad1,Kim}.

In the case where all the families contain at most two members, the primary constraints will have vanishing Poisson brackets
with all of the constraints
%%%%%%
\footnote{This can be easily verified by noticing 
that in this case the canonical Hamiltonian can always be written in the form 
$H_c(q,p,\xi) = H_0(q,p)+\xi^\alpha T_\alpha$ \cite{Batlle}, where the Lagrange
multipliers $\xi^\alpha$ are the variables conjugate to the primary constraints,
and implement the secondary constraints $T_\alpha \approx 0$.},
%%%%%%
amounting to a vanishing of the last term in (\ref{parameter}).
 In that case we can also relax the 
above assumption concerning the constancy of the structure functions
$V^{ab}_{ij}$, since our iterative scheme already terminates with
equation (\ref{startingpoint}) with $N_a=N_b=1$, and we have for the generator
\be\label{generalsol}
G = \sum_{a=1}^M \left[\left(\frac{d\alpha^a}{dt} - 
\sum_{b=1}^M \alpha^b V_{11}^{ba}\right)\phi^{(a)}_0
+ \alpha^a\phi^{(a)}_1\right]\;.
\ee
The following
modified version of the Nambu-Goto model has these features.

Consider the Lagrangian \cite{Batlle}
\be\label{NGlagrangian}
L = \int d\sigma \left(\frac{1}{2}\frac{{\dot x}^2}{\lambda}
- \frac{\mu}{\lambda}{\dot x}x' + \frac{1}{2}\frac{\mu^2}{\lambda} x'^2 - \frac{1}{2}\lambda x'^2 \right)\;,
\ee
where the 4-vector $x^\mu(\tau,\sigma)$ labels the coordinates
of a ``string" parametrized by $\tau$ and $\sigma$,
with the ``dot" and ``prime" denoting the derivative with respect
to $\tau$ and $\sigma$, respectively.
There are two primary constraints, $\pi_1 \approx 0$ and $\pi_2 \approx 0$, where
$\pi_1$ and $\pi_2$ are the momenta conjugate 
to the fields $\lambda(\tau,\sigma)$ and $\mu(\tau,\sigma)$, respectively. 
Hence in our notation
\be\label{primary}
\phi_0^{(1)} = \pi_1 \;,\quad \phi_0^{(2)} = \pi_2 \;.
\ee
The canonical Hamiltonian reads,
\be\label{NGhamiltonian}
H_c = \int d\sigma \{\frac{\lambda}{2}(p^2 + x'^2) + \mu p\cdot x' \}\;,
\ee
where $p_\mu$ is the four-momentum conjugate to the coordinate $x^\mu$.
The conservation in time of the primary constraints leads respectively
to secondary constraints, which in our notation read
\be\label{secondaryconstr}
\phi^{(1)}_1 = \frac{1}{2}(p^2 + x'^2) \approx 0 \;,
\quad \phi^{(2)}_1 = p\cdot x' \approx 0\;.
\ee
One readily checks that there are no further constraints.

We see that the secondary constraints are just the primary constraints of the
Nambu-Goto string model. They satisfy the familiar Poisson brackets
%%%%%%
\footnote{We suppress the $\tau$ variable.}
%%%%%
\be
[\phi^{(1)}_1(\sigma),\phi^{(1)}_1({\sigma}')] =
 \phi^{(2)}_1 (\sigma)\partial_\sigma \delta(\sigma-{\sigma}')
-  \phi^{(2)}_1 ({\sigma}')\partial_{{\sigma}'} \delta(\sigma-{\sigma}')
\ee
\be
[\phi^{(1)}_1(\sigma),\phi^{(2)}_1({\sigma}')] =
 \phi^{(1)}_1 (\sigma)\partial_\sigma \delta(\sigma-{\sigma}')
-  \phi^{(1)}_1 ({\sigma}')\partial_{{\sigma}'} \delta(\sigma-{\sigma}')
\label{NGalgebra}
\ee
\be
[\phi^{(2)}_1(\sigma),\phi^{(2)}_1({\sigma}')] =
 \phi^{(2)}_1 (\sigma)\partial_\sigma \delta(\sigma-{\sigma}')
-  \phi^{(2)}_1 ({\sigma}')\partial_{\sigma'} \delta(\sigma-{\sigma}')\;.
\ee
All other Poisson brackets vanish. The constraints are 
seen to be first class. In our terminology, we thus have two families, 
each with two members.

The canonical Hamiltonian is of the form
\be\label{canonHamiltonian}
H_c = \int d\sigma (\lambda\pi^{(1)}_1(\sigma) + \mu \phi^{(2)}_1(\sigma))\;.
\ee
The structure functions $V_{ij}^{ab}$ are read off from the
Poisson brackets
\be
[H_c,\phi^{(1)}_1] = -\lambda\partial_\sigma\phi^{(2)}_1
-2\lambda' \phi^{(2)}_1 - \mu\partial_\sigma\phi^{(1)}_1 
- 2\mu' \phi^{(1)}_1\;,\label{HCcommutator1}
\ee
\be
[H_c,\phi^{(2)}_1] = -\lambda\partial_\sigma\phi^{(1)}_1
-2\lambda' \phi^{(1)}_1 - \mu\partial_\sigma\phi^{(2)}_1 
- 2\mu' \phi^{(2)}_1\;,\label{HCcommutator2}
\ee
to be
\bear
V_{11}^{11}(\sigma,\sigma') &=& V_{11}^{22}(\sigma,\sigma') =
-\left(\mu(\sigma)\partial_\sigma + 
2\mu'(\sigma)\right)\delta(\sigma-\sigma')\;,
\nonumber\\
V_{11}^{12}(\sigma,\sigma') &=& V_{11}^{21}(\sigma,\sigma') =
-\left(\lambda(\sigma)\partial_\sigma + 2\lambda'(\sigma)\right)\delta(\sigma-\sigma')\;.\label{structurefunctions}
\eear
Since for the example in question $N_1=N_2=1$, it follows from
(\ref{startingpoint}), that our iterative scheme for finding the solution
already ends at the first step, with
$\epsilon_0^{(a)}$ given by
\be
\epsilon_0^{(a)} = \frac{d\alpha^a}{d\tau} - \int d\sigma' \sum_{b=1}^{2}\alpha^b(\sigma')V_{11}^{ba}(\sigma',\sigma)\;.\nonumber
\ee
We thus obtain
\bear
\epsilon_0^{(1)} &=& \frac{d\alpha^1}{d\tau}
-\mu\partial_\sigma\alpha^1 + \mu'\alpha^1
-\lambda\partial_\sigma\alpha^2 + \lambda'\alpha^2 \;,\label{epsilon1}\\
\epsilon_0^{(2)} &=& \frac{d\alpha^2}{d\tau}
-\mu\partial_\sigma\alpha^2 + \mu'\alpha^2
-\lambda\partial_\sigma\alpha^1 + \lambda'\alpha^1 \label{epsilon1}\;.
\eear
From (\ref{generator}) and (\ref{delta}) we now compute the corresponding
transformation laws for the fields to be
\bear
\delta x^\mu &=& \alpha^1 p^\mu + \alpha^2 \partial_\sigma x^\mu \;,\nonumber\\
\delta \lambda &=& \epsilon_0^{(1)}\;,\quad \delta \mu = \epsilon_0^{(2)}\;.
\label{gaugetransformations}
\eear
Making use of the expressions for $\epsilon^{(a)}_0$ derived above, we verify that our results
(\ref{gaugetransformations}) agree with that quoted in the literature
\cite{Batlle}.

To summarize, we have shown that the equations defining the restrictions to be imposed on
the gauge parameters in (\ref{generator}) could be solved following a simple
iterative scheme, in the case where the structure functions $C_{a_1b}^{c_2}$
in eq. (\ref{parameter})
vanish and $V_a^{b_2}$ are constants. We have then applied 
the general ideas to the case of a non-trivial model with a two-family
constraint structure, sharing some
properties with the familiar Nambu-Goto model of string theory. Since
each family of constraints consisted only of two members, our general solution was applicable, although the structure functions $V_a^{b_2}$ 
are functions of the fields. We thereby recovered the local symmetry 
transformations quoted in the literature.

\section*{Acknowledgement}

One of the authors (R.B.) would like to thank the Alexander von Humboldt Foundation for providing financial support making this collaboration
possible.

%%%%%%%%%%%%%%%%%%%%%%%%%%%%%%%%%%%%%%


\begin{thebibliography}{999}
%%%%%%%%%%%%%%%%%%%%%%%%%%%%%%%%%%%%%
\bibitem{Mukunda} E.C.G. Sudarshan and N. Mukunda, {\it Classical Dynamics:
a Modern Perspective}, Wiley, New York, 1974. 
\bibitem{Gitman} D.M. Gitman and I.V. Tyutin, {\it Quantization of
fields with constraints}, Springer-Verlag, Heidelberg, 1990.
\bibitem{Chaichian1} M. Chaichian and D. Louis Martinez, J. Math. Phys.
{\bf 35} (1994) 6536.
\bibitem{Shirzad1} A. Shirzad, J. Phys. {\bf A31} (1998) 2747.
\bibitem{Girotti}M.E.V. Costa, H.O. Girotti and T.J.M. Simoes,
Phys. Rev. {\bf D32} (1985) 405.
\bibitem{Henneaux} M. Henneaux, C. Teitelboim and J. Zanelli,
Nucl. Phys. {\bf B332} (1990) 169.
\bibitem{Chaichian} A. Cabo, M. Chaichian and D. Louis Martinez,
J. Math. Phys {\bf 34} (1993) 5646.
\bibitem{Wipf} V. Mukhanov and A. Wipf, Int. J. Mod. Phys. {\bf A10} (1995) 579.
\bibitem{Banerjee1} R. Banerjee, H.J. Rothe and K.D. Rothe, 
Heidelberg preprint HD-THEP-99-17, hep-th/9906072, to appear in Phys. Lett. B.
\bibitem{Banerjee2} R. Banerjee, H.J. Rothe and K.D. Rothe,
Heidelberg preprint HD-THEP-99-24, hep-th/9907217. 
\bibitem{Kim}For a non-trivial example see Y.-W. Kim and K.D. Rothe,
Int. J. Mod. Phys. {\bf A13} (1998) 4183. 
\bibitem{Batlle}C. Batlle, J. Gomis, J. Paris and J. Roca,
Nucl. Phys. {\bf B329} (1990) 139.
\bibitem{Dirac} P.A.M. Dirac, Can. J. Math. {\bf 2} (1950) 129; 
{\it Lectures on Quantum Mechanics}, Yeshiva University, 1964.
\bibitem{Shirzad2}A. Shirzad and M. Shabani Moghadam, IUT-Phys/98-25,
February 1999.

\end{thebibliography}
\end{document}